\documentclass[global,twocolumn]{svjour}
\usepackage{graphics}
\usepackage{graphicx}
\journalname{}

\begin{document}

\title{Dark resonances for ground state transfer of molecular quantum gases}
\author{Manfred J. Mark\inst{1}, Johann G. Danzl\inst{1}, Elmar Haller\inst{1}, Mattias Gustavsson\inst{1}, Nadia Bouloufa\inst{2}, Olivier Dulieu\inst{2}, Houssam Salami\inst{3}, Tom Bergeman\inst{3}, Helmut Ritsch\inst{4}, Russell Hart\inst{1} \and Hanns-Christoph N\"agerl\inst{1}}

\authorrunning{M.J. Mark {\em et al.}}
\mail{johann.danzl@uibk.ac.at}

\institute{Institut f{\"u}r Experimentalphysik und Zentrum f\"{u}r Quantenphysik, Universit{\"a}t Innsbruck, Technikerstra{\ss}e 25, A--6020 Innsbruck, Austria \and Laboratoire Aim\'e Cotton, CNRS, Universit\'e Paris-Sud, B\^{a}t. 505, 91405 Orsay Cedex, France \and Department of Physics and Astronomy, SUNY Stony Brook, NY 11794-3800, USA \and Institut f\"ur Theoretische Physik und Zentrum f\"{u}r Quantenphysik, Universit{\"a}t Innsbruck, Technikerstra{\ss}e 25, A--6020 Innsbruck, Austria}

\date{Received: date / Revised version: date}

\maketitle

\begin{abstract}
One possible way to produce ultracold, high-phase-space-density quantum gases of molecules in the rovibronic ground state is given by molecule association from quantum-degenerate atomic gases on a Feshbach resonance and subsequent coherent optical multi-photon transfer into the rovibronic ground state. In ultracold samples of Cs$_2$ molecules, we observe two-photon dark resonances that connect the intermediate rovibrational level $|v\!=\!73, J\!=\!2>$ with the rovibrational ground state $|v\!=\!0,  J\!=\!0>$ of the singlet $X^1\Sigma_g^+$ ground state potential. For precise dark resonance spectroscopy we exploit the fact that it is possible to efficiently populate the level $|v\!=\!73,  J\!=\!2>$ by two-photon transfer from the dissociation threshold with the stimulated Raman adiabatic passage (STIRAP) technique. We find that at least one of the two-photon resonances is sufficiently strong to allow future implementation of coherent STIRAP transfer of a molecular quantum gas to the rovibrational ground state $|v\!=\!0,  J\!=\!0>$.
\end{abstract}
\maketitle

\section{Introduction}
\label{intro}Laser cooling of atoms and the production of quantum degenerate atomic Bose and Fermi gases have revolutionized the field of atomic physics \cite{Southwell2002}. For molecular systems, ultralow temperatures and high phase space densities are much more difficult to achieve. Laser cooling of molecules has not yet been demonstrated, and with alternative cooling and slowing techniques such as buffer gas cooling and Zeeman slowing high phase space densities are yet out of reach \cite{Doyle2004,Krems2008,Meerakker2008}. In photoassociation experiments from magneto-optical traps, \cite{Jones2006,Nikolov2000,Sage2005,Viteau2008,Deiglmayr2008}, cold samples of deeply bound molecules in the lowest vibrational levels have been created. Yet, the phase space densities are far away from the quantum degenerate regime. In the limit of extremely weak binding, molecular Bose-Einstein condensation could be achieved \cite{Fermi2008} by using the trick of first cooling an atomic Fermi gas to high phase space densities and subsequently associating pairs of atoms to molecules. For molecules composed of Fer\-mions, collisional stability of the highly excited molecules is assured as a result of a Pauli blocking effect. Here, we are interested in ultracold and dense molecular systems in specific deeply bound rovibrational levels. Such samples are of high interest for fundamental studies in physics and chemistry, ranging from ultracold chemistry \cite{Krems2005} and few-body collisional physics \cite{Staanum2006,Zahzam2006} to high resolution spectroscopy \cite{Zelevinsky2008,DeMille2008}, to applications in quantum processing \cite{DeMille2002}, and to the formation of dipolar quantum gases and dipolar Bose-Einstein condensates \cite{Goral2002,Baranov2002}. For these experiments full control over the molecular wave function is desired. In addition, high densities are required for molecular quantum gas studies. Only in the rovibronic ground state, i.e. the lowest energy level of the electronic ground state, is collisional stability assured.

For the production of molecular quantum gases in the absolute ground state, we follow a scheme in which the technique of stimulated two-photon transfer is repeatedly applied to molecules associated on a Feshbach resonance from a high-density sample of ultracold atoms such as a Bose-Einstein condensate (BEC). The initially very loosely bound molecules are to be transferred in a few successive steps to the rovibrational ground state, acquiring more and more binding energy.
The scheme has several advantages. It is fully coherent, not relying on spontaneous processes, allowing high state selectivity, and it involves only a comparatively small number of intermediate levels. The scheme is expected to allow the removal of a ground state binding energy of typically 0.5 eV for an alkali dimer without appreciably heating the molecular sample. It essentially preserves phase space density and coherence of the particle wave function, allowing the molecular sample to inherit the high initial phase space density from the atomic sample. Ideally, the scheme will ultimately result in the formation of a molecular BEC. A major challenge is given by the low radial wave function overlap between successive molecular levels, potentially leading to prohibitively low transition rates for the two-photon transitions that could only be compensated by the use of further (smaller) transfer steps.

In a crucial experiment, Winkler {\it et al.} \cite{Winkler2007} demonstrated that coherent two-photon transfer by means of the stimulated Raman adiabatic passage (STIRAP) technique \cite{Bergmann1998} can efficiently be implemented with quantum gases of weakly bound Feshbach molecules. In this work, the transferred molecules, in this case Rb$_2$, were still weakly bound with a binding energy of much less than $10^{-4}$ of the binding energy of the rovibrational ground state. In particular, wave function overlap of the final level with the rovibrational ground state is negligible. Nevertheless, an important result of this experiment was the demonstration that, even with excitation near the excited S+P asymptote, parasitic excitation of unwanted molecular transitions by the STIRAP laser beams could largely be avoided. Recently, Danzl {\it et al.} \cite{Danzl2008} showed efficient coherent STIRAP transfer into deeply bound rovibrational levels in the quantum gas regime. More specifically, transfer into the rovibrational level $|v\!=\!73, J\!=\!2>$ of the singlet $X^1\Sigma_g^+$ molecular potential of the Cs dimer was demonstrated. This level is bound by 1061 wavenumbers, more than one-fourth of the binding energy of the rovibrational ground state. Here, as usual, $v$ and $J$ denote the vibrational and rotational quantum numbers, respectively. This intermediate level was chosen as to give a balanced distribution for the wave function overlap in a four-photon transfer scheme to the ground state, i.e. to assure that all four dipole transition moments are of comparable magnitude. This level could thus serve as a transfer state towards the rovibrational ground state $|v\!=\!0, J\!=\!0>$, allowing coherent ground state transfer with two two-photon transitions. Also recently, Ni {\it et al.} \cite{Nie2008} could demonstrate transfer all the way into the rovibrational ground state $|v\!=\!0, J\!=\!0>$ of the singlet $X^1\Sigma^+$ molecular potential in a quantum gas of KRb molecules. The transfer could be achieved in a single step as a result of the favorable run of the excited state potentials in the case of heteronuclear alkali dimers \cite{Stwalley2004}. Also, the lowest rovibrational level of the Rb$_2$ triplet $a^3\Sigma_u^+$ potential could recently be populated in the quantum gas regime using the STIRAP technique \cite{Lang2008}.

Here, in an ultracold and dense sample of Cs mole\-cules, we present two-photon dark resonances connecting the rovibrational level $|v\!=\!73, J\!=\!2>$ of the Cs dimer singlet $X^1\Sigma_g^+$ molecular potential with the rovibrational ground state $|v\!=\!0, J\!=\!0>$. Starting from $|v\!=\!73, J\!=\!2>$, we first perform molecular loss spectroscopy by laser excitation in the wavelength range from 1329 nm to 1365 nm to search for and identify suitable excited state levels of the mixed $(A^1\Sigma_u^+ - b^3\Pi_{0u}) \ 0_u^+$ excited molecular potentials. These levels are 9893 to 10091 wavenumbers above the rovibronic ground state, corresponding to a wavelength range from 1011 nm to 991 nm for the transition to the rovibronic ground state. We then perform dark state spectroscopy by simultaneous laser irradiation near 1350 nm and 1000 nm. We find several dark resonances, from which we derive normalized transition strengths and find that at least one of the two-photon transitions is favorable for ground state transfer.

\section{Molecular energy levels and laser transitions}
\label{sec:2}

\begin{figure}[t]
\resizebox{0.45\textwidth}{!}{ \includegraphics{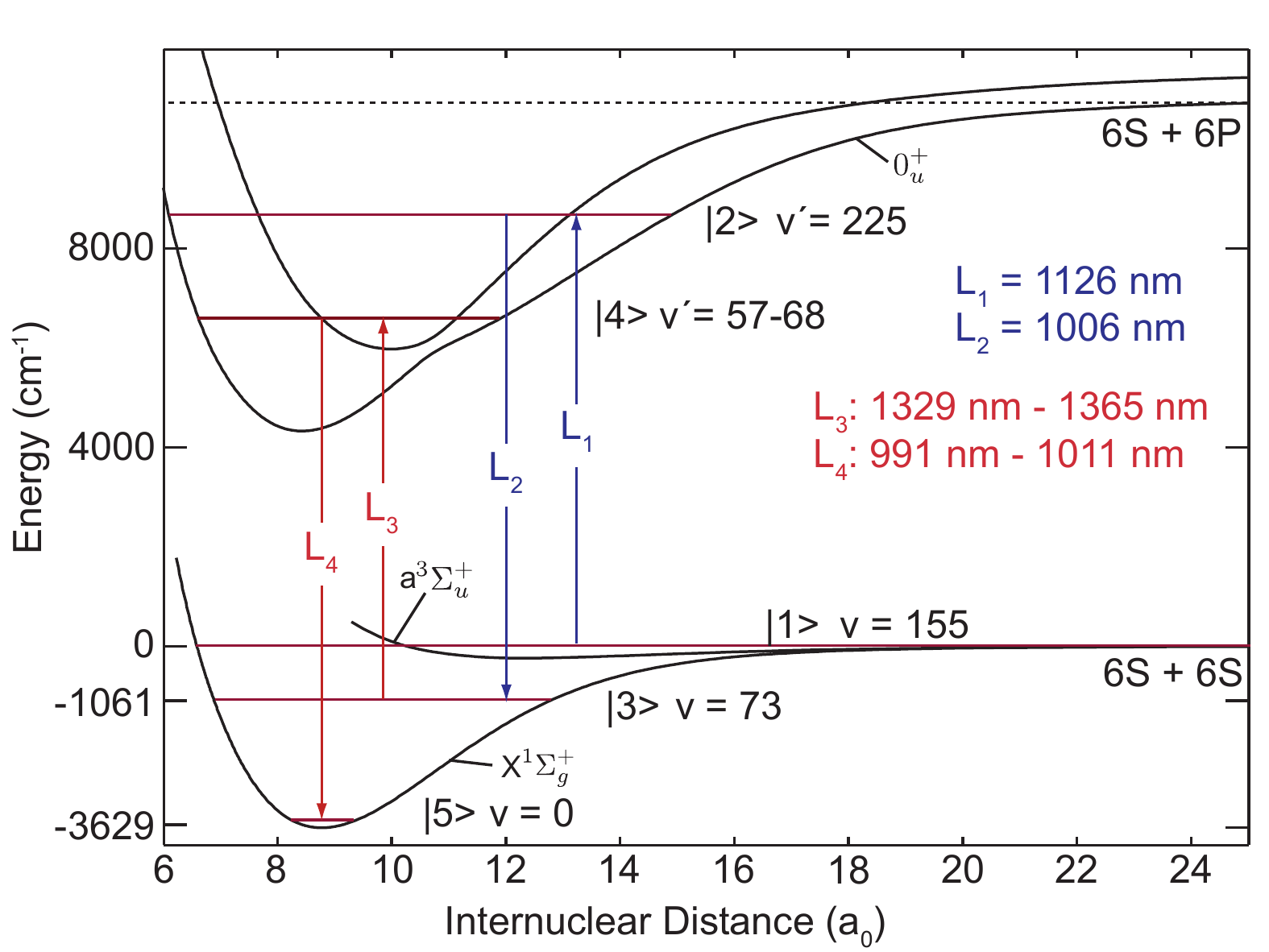} }
\caption{Molecular level scheme for Cs$_2$. Molecules in a weakly
bound Feshbach level $|1\!\!>= |v\!\approx\!155>$ (not resolved
near the 6S+6S asymptote) are transferred to the rovibrational
level $|3\!\!>=|v\!=\!73,J\!=\!2\!\!>$ of the singlet
$X^1\Sigma_g^+$ ground state potential with a binding energy of
$1061$ cm$^{-1}$ by a two-photon STIRAP process \cite{Danzl2008}
involving lasers $L_1$ and $L_2$ near 1126 nm and 1006 nm. The
following two-photon transition from $|3\!\!>$ to
$|5\!\!>=|v\!=\!0,J\!=\!0\!\!>$ and also to
$|v\!=\!0,J\!=\!2\!\!>$ is then probed by lasers $L_3$ and $L_4$
near 1350 nm and 1000 nm, respectively. Level $|2\!\!>$ is the
225th level of the electronically excited coupled $ (A^1\Sigma_u^+
- b^3\Pi_{0u}) \ 0_u^+$ potentials. Here, we probe suitable
candidate levels for $|4\!\!>$, connecting $|3\!\!>$ to $|5\!\!>$.
These candidate levels also belong to the $0_{u}^{+}$ coupled
state system and include levels with coupled channel vibrational
numbers $v'\!=\!57$ to $68$. The position of the vertical arrows
is not meant to reflect the internuclear distance at which the
transition takes place.}
\label{fig:1}       
\end{figure}

\begin{figure}[t]
\resizebox{0.45\textwidth}{!}{ \includegraphics{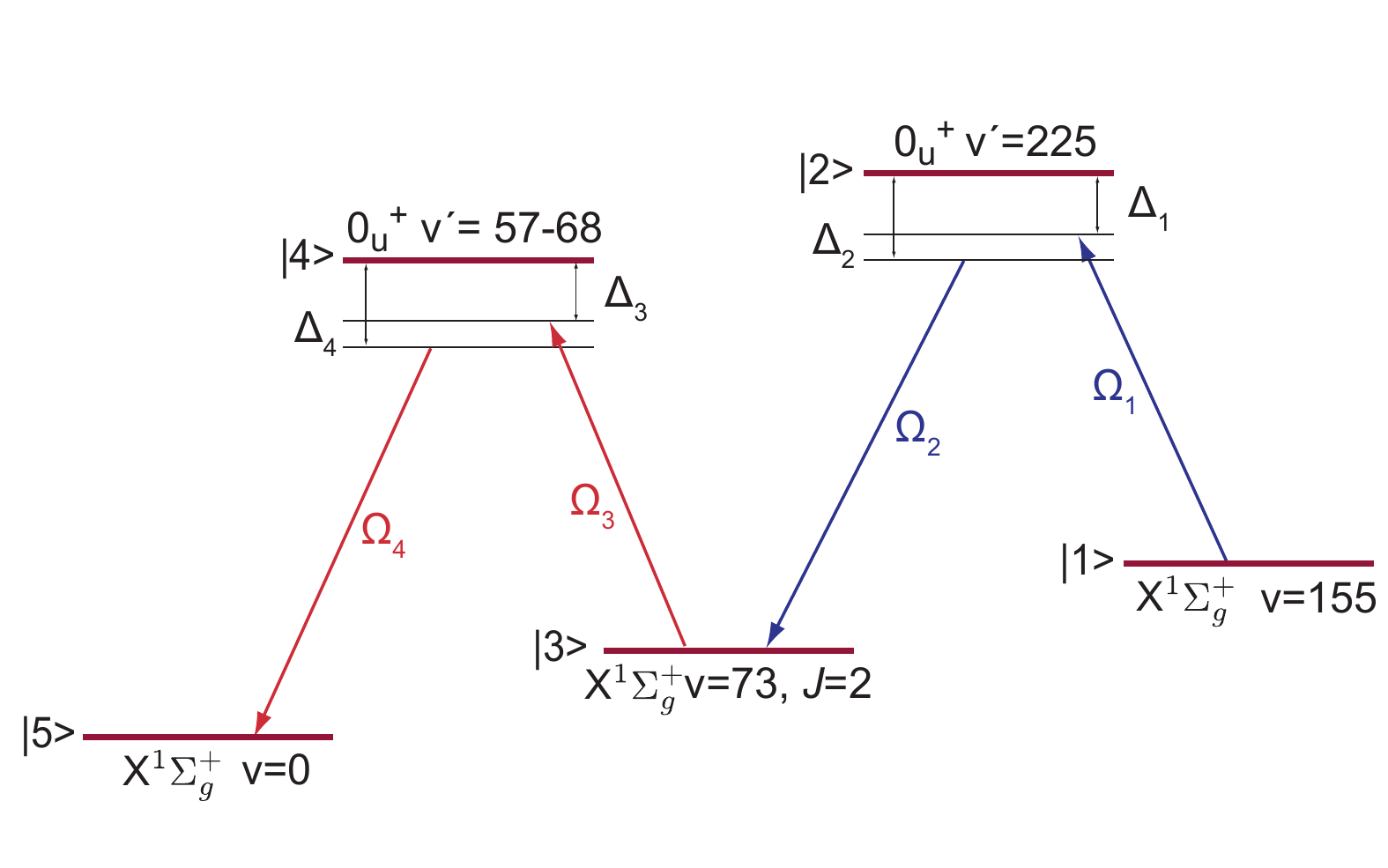} }
\caption{5-level distorted M-scheme. The one-photon-detunings and
Rabi frequencies of $ L_i $ are $\Delta_i$ and $\Omega_i$,
$i=1,2,3,4$. For STIRAP to $ |v\!=\!73, J\!=\!2>$ the detunings
for $L_1$ and $L_2$ are $\Delta_1\approx0\approx\Delta_2$.}
\label{fig:2}       
\end{figure}

Fig.\ref{fig:1} shows the energy of the relevant Cs$_2$ molecular states and the optical transitions for our transfer scheme. State $ |1\!\!> $ is the initial weakly bound Feshbach state that we populate out of an atomic BEC of Cs atoms via Feshbach association \cite{Herbig2003}. For the transfer from $ |1\!\!> $ to the ro-vibrational ground state $ |5\!\!> = |v\!=\!0, J\!=\!0>$, three intermediate levels $ |2\!\!>, |3\!\!>  $, and $ |4\!\!> $ are needed. All five molecular levels are coupled by two two-photon transitions in a distorted M-shaped configuration as shown in Fig.\ref{fig:2}. Levels $ |2\!\!> $ and $ |4\!\!>  $ belong to the excited mixed $(A^1\Sigma_u^+ - b^3\Pi_{0u}) \ 0_u^+$ potentials. We have identified level $ |2\!\!> $ as the 225th one of the coupled $ 0_u^+ $ system, with an uncertainty of $2$ in the absolute numbering, and $ |3\!\!> $ is the level with $v\!=\!73$ and $J\!=\!2$ of the $X^1\Sigma_g^+$ ground state potential \cite{Danzl2008}. A two-photon laser transition with laser $L_1$ at $1126$ nm and laser $L_2$ at $1006$ nm couples $ |1\!\!> $ to $ |3\!\!> $ via $ |2\!\!> $. There are now several possibilities for coupling  $ |3\!\!> $ to $ |5\!\!> $, differing in the choice of the excited state $ |4\!\!> $. The aim of this work is to identify a suitable state $ |4\!\!> $ from the $(A^1\Sigma_u^+ - b^3\Pi_{0u}) \ 0_u^+$ potentials with sufficient wave function overlap with both $ |3\!\!> $ and $ |5\!\!> $. We search for state $ |4\!\!> $ in the energy range of $9893$ to $10091$ wavenumbers above the rovibrational ground state $ |5\!\!> $. Molecular structure calculations as outlined in Sec. \ref{sec:4} show that in this range there are candidate states for $ |4\!\!> $ that have dipole transition matrix elements with both $ |3\!\!> $ and $ |5\!\!> $ of comparable magnitude, allowing optimum STIRAP performance. The wavelengths for the lasers $L_3$ and $L_4$ driving the associated two-photon transition are near 1350 nm and 1000 nm, respectively. We derive all laser light for driving the molecular transitions from highly stable, widely tunable diode laser systems with kHz linewidths. For short term stability, the lasers are all locked to narrow-band optical resonators. For long term stability, the optical resonators are referenced to an infrared, fiber-laser-based frequency comb, covering the wavelength range from about 980 nm to about 2000 nm.

\section{Preparation of a molecular quantum gas in $v\!=\!73, J\!=\!2$}
\label{sec:3}Our sample preparation procedure follows Ref. \cite{Danzl2008}. In summary, we first produce a cigar-shaped BEC of typically $1.5\times10^5 $ cesium atoms in the lowest hyperfine sublevel $F\!=\!3, \ m_F\!=\!3$ in a crossed optical dipole trap. As usual, $F$ is the atomic angular momentum quantum number, and $m_F$ its projection. The trapping light at 1064.5 nm is derived from a single-frequency, highly-stable Nd:YAG laser. Using a $d$-wave Feshbach resonance at 4.8 mT \cite{Mark2007} we then produce a quantum gas of weakly bound Feshbach molecules out of the BEC \cite{Herbig2003}. For this, we first ramp the magnetic field from the BEC production value of 2.0 mT to 4.9 mT, slightly above the Feshbach resonance. The molecules are produced on a downward sweep at a typical sweep rate of 0.025 mT/ms. The resulting ultracold sample contains up to 11000 molecules, immersed in the bath of the remaining BEC atoms. For the present experiments we shut off the trap and perform all subsequent measurements in free flight. This reduces the particle density, in particular during the later detection stage of the experiment, and hence reduces atom-molecule collisional loss, thus increasing the molecular signal. Following two avoided state crossings while further sweeping the magnetic field to lower values, we transfer the molecules via a weakly bound, open channel $s$-wave molecular state into the still weakly bound, closed channel $s$-wave molecular state $ |1\!\!> $ by magnetic field ramping \cite{Danzl2008}. This is the starting state for the subsequent optical transfer. As with all other weakly bound Feshbach states, it belongs to both the $X^1\Sigma_g^+$ ground state potential and the lowest triplet $a^3\Sigma_u^+$ potential and is hence of mixed character. It has zero rotational angular momentum. At a field of 1.9 mT, it has a binding energy of $ 5 $ MHz$\times h$, where $h$ is Planck's constant, with respect to the $F\!=\!3, m_F\!=\!3$ two-atom asymptote \cite{Mark2007}. We detect molecules in $|1\!\!>$ by reverse magnetic field ramping, leading to dissociation on the Feshbach resonance at 4.8 mT, and by subsequent imaging of the resulting atoms \cite{Herbig2003}.%

We transfer the molecules from $|1\!\!>$ to the rovibrational level $|3\!\!> = |v\!=\!73, J\!=\!2>$ with the STIRAP technique \cite{Danzl2008}. For this, about 3 ms after molecule production, with the magnetic field ramping completed, laser $L_2$ at 1006 nm is pulsed on first and then laser $L_1$ at 1126 nm. Both lasers are on resonance within a few kHz. The pulse overlap time is about $10 \ \mu$s. With peak Rabi frequencies of $\Omega_1\!\approx\!2\pi\!\times\!3 $ MHz and $\Omega_2\!\approx\!2\pi\!\times\!6$ MHz we transfer about 80 \% of the molecules to $|3\!\!>$. We find that the molecular sample is not heated as a result of the STIRAP transfer. A residual kinetic energy on the order of k$_B \times 10$ nK comes from the expansion energy of the initial atomic sample. Our current procedure allows us to produce a sample of up to 8000 molecules in state $|3\!\!>$ every 12 s. For the loss spectroscopy as detailed below, we irradiate the molecules in $|3\!\!>$ with light near 1350 nm for a certain waiting time. We then measure the fraction of molecules that have remained in $|3\!\!>$. For this, we transfer the remaining molecules back to $|1\!\!>$ using the reverse STIRAP process and determine the number of molecules in $|1\!\!>$. Without irradiation with light near 1350 nm we transfer more than 65\% of the molecules from $|1\!\!>$ to $|3\!\!>$ and back to $|1\!\!>$ \cite{Danzl2008}.

\section{Loss spectroscopy}
\label{sec:4}

\begin{figure}[t]
\resizebox{0.45\textwidth}{!}{ \includegraphics{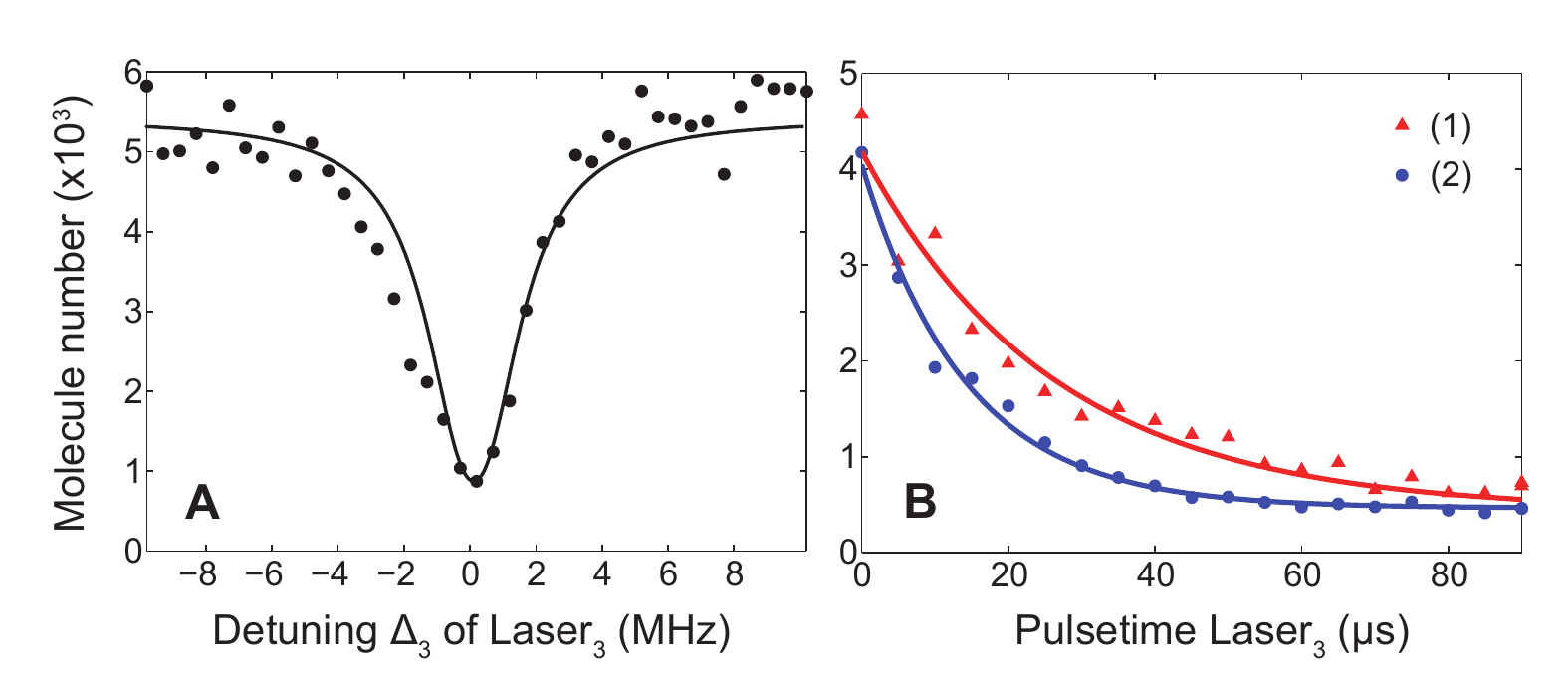} }
\caption{Loss resonances for excitation near 1351 nm from
$|3\!\!>=|v\!=\!73, J\!=\!2>$ of the $X^1\Sigma_g^+$ ground state
potential. \textbf{(A)} Loss of molecules in $|3\!\!>$ as a
function of laser detuning $\Delta_3$ near 1351 nm after a waiting
time of $20\,\mu$s. The solid line represents a model calculation
matched to the data yielding an excited state natural linewidth of
$2 \pi \times 2$ MHz. \textbf{(B)} Time dependence of molecular
loss on resonance at 1351 nm for two different laser intensities.
(1) 270 $\pm$ 80 mW/cm$^2$, (2) 570 $\pm$ 80 mW/cm$^2$. The fitted
exponential decay gives the decay constants $\tau = 26\pm4\ \mu$s
for 270 mW/cm$^2$ and $\tau = 14\pm2\ \mu$s for 570 mW/cm$^2$.}
\label{fig:3}
\end{figure}

Prior to the present experiments, the energies of the levels with predominant $A^{1}\Sigma_{u}^{+}$ character in the region of interest were established to about $\pm$ 0.06 cm$^{-1}$ by fits \cite{Salami2008} to data obtained by Fourier transform spectroscopy (FTS) at Laboratoire Aim\'{e} Cotton (LAC) using transitions to the $X ^{1}\Sigma_{g}^{+}$ state. However, the predominantly $b^{3}\Pi_{0u}$ levels were only known to about $\pm$ 2 cm$^{-1}$ because this region was above that for which data was obtained from $2 ^{3}\Delta_{1g} \rightarrow b ^{3}\Pi_{0u}$ emission \cite{Xie2008}, but lower than the regime where $b^{3}\Pi_{0u}$ levels acquire sufficient singlet character (by spin-orbit mixing) to be observed in the FTS work. Paradoxically, the predominantly $b ^{3}\Pi_{0u}$ levels are of special interest here because they happen to have significant singlet character over regions of the internuclear distance that are most important for transitions of interest in this work.

The coupled channel calculations used to characterize the level structure
of the strongly interacting $A ^{1}\Sigma_{u}^{+}$ and $b
^{3}\Pi_{0u}$ states employed methods developed from previous work
on $A$ and $b$ states of K$_{2}$ \cite{LisK2,MRMK2}, RbCs
\cite{TB03}, Na$_{2}$ \cite{QiNa2}, and Rb$_{2}$ \cite{HS08}.  The
DVR approach \cite{CM} was used to calculate eigenvalues primarily
for two coupled channels, although some information on $b
^{3}\Pi_{1u}$ was found in the FTS data from LAC. Similar
computational approaches, differing in the detailed numerical
methods, have been applied recently also to the $A$ and $b$ states
of NaRb \cite{FerbNaRb}.

Because of the initial $\pm$ 2 cm$^{-1}$ uncertainty in the
positions of $b ^{3}\Pi_{0u}$ levels of interest, we decided to perform a systematic, broad-range search around expected transition energies in the wavelength range from 1329 nm to 1365 nm. For this, we perform double STIRAP from $|1\!\!>$ to $|3\!\!>$ and back with a waiting time of typically $\tau = 1$ ms. During the waiting time, we irradiate the sample with laser $L_3$ at an estimated intensity of $5\cdot10^4$ mW/cm$^2$. Laser $L_3$ is a diode laser with grating feedback. On the timescale of our experiment, the resonator of the laser is sufficiently stable, allowing systematic tuning of the laser without locking the laser to its external resonator. We step the laser frequency in units of typically $20$ MHz by tuning the piezo element on the grating. We monitor the laser wavelength with a home-built wavemeter at approximately 300 MHz accuracy. For the initial broad range line search we increased the repetition rate of the experiment by stopping evaporative cooling slightly before condensation sets in. While stepping the laser, taking data points essentially at the cycle rate corresponding to the sample production time, we look for a dip in the molecule number. Once such a dip is found, typically consisting of a few data points, we perform a more precise scan by locking the laser to the external, highly-stable resonator and then the external resonator to the infrared frequency comb. This allows us to detune the laser with kHz precision. Fig.\ref{fig:3} (A) shows a typical loss resonance near 1351 nm. We reduce the laser intensity such that on resonance at most $80\%$ of the molecules are lost within $20\ \mu$s.
From such measurements the transition strength as given by the normalized Rabi frequency and the natural linewidth of the excited state can be deduced.
The typical width of the excited state molecular levels that we have identified is $2 \pi \times 2$ MHz, in agreement with typical expected lifetimes. Fig.\ref{fig:3} (B) shows a measurement of the time dependence of the molecular loss. Here, we step the waiting time $\tau$ from $0$ to $50 \ \mu$s, while the laser is kept on resonance. In total, we have found 7 excited levels belonging to the $(A^1\Sigma_u^+ - b^3\Pi_{0u}) \ 0_u^+$ coupled state system. They are listed in Table \ref{tab:1} along with the dominant overall character (either $A^1\Sigma_u^+$ state or $b^3\Pi_{0u}$ state) of the vibrational wave function as determined from the coupled state calculations. Within the wavelength range from 1329 nm to 1365 nm, theory predicts the existence of 5 more states of the  $0_u^+$ coupled state system, whose energies are also displayed in Table \ref{tab:1}. For most of them, the wave function overlap is not expected to be favorable for STIRAP transfer to X $^1\Sigma_g^+$ $|v\!=\!0>$. However, an improved model of the energy level structure, based on all the data except one FTS point with a large residual, fits the observed transitions to a rms residual error of 0.02 cm$^{-1}$, indicating that additional resonances can be found with searches over very limited ranges of laser frequency.

\section{Dark resonances with $|v\!=\!0, J\!=\!0>$ and $|v\!=\!0, J\!=\!2>$ }
\label{sec:5}
\begin{figure}[t]
\resizebox{0.45\textwidth}{!}{ \includegraphics{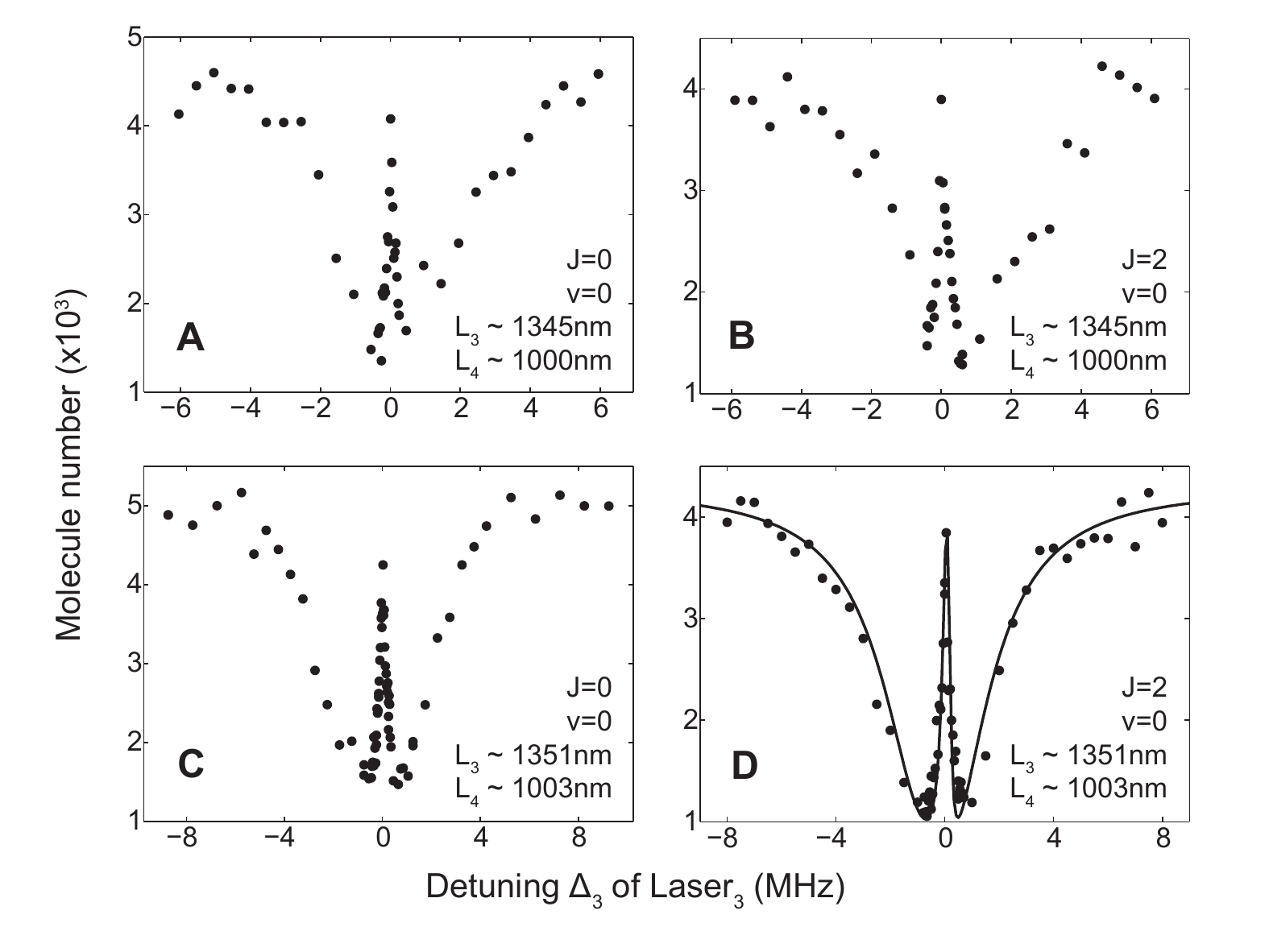} }
\caption{Dark resonances involving $X^1\Sigma_g^+$  state levels
$|v\!=\!73, J\!=\!2>$ and $|v\!=\!0>$ for two different
intermediate levels. \textbf{(A} and \textbf{B)} Dark resonances
with $X^1\Sigma_g^+$ $|v\!=\!0, J\!=\!0>$ and $|v\!=\!0, J\!=\!2>$
involving the $0_u^+$ excited state level $|v'\!=\!63, J\!=\!1>$
at an excitation wavelength near 1345 nm. \textbf{(C} and
\textbf{D)} Dark resonances with $X^1\Sigma_g^+$ $|v\!=\!0,
J\!=\!0>$ and $|v\!=\!0, J\!=\!2>$ involving the excited state
level $|v'\!=\!61, J\!=\!1>$ at an excitation wavelength near 1351
nm. The solid line in \textbf{(D)} is the result of a model
calculation, solving the three-level master equation including
laser bandwidth and loss, matched to the data giving
$\Omega_3\!=\!2\pi\!\times\!6$ kHz $ \sqrt{I/(\mathrm{mW/cm}^2)}$
and $\Omega_4\!=\!2\pi\!\times\!4$ kHz $
\sqrt{I/(\mathrm{mW/cm}^2)}$ for $X^1\Sigma_g^+$ $|v=0, J=2>$. The
corresponding calculation for $X^1\Sigma_g^+$ $|v=0, J=0>$ yields
$\!2\pi\!\times\!5$ kHz $ \sqrt{I/(\mathrm{mW/cm}^2)}$.}
\label{fig:4}       
\end{figure}

In our recent work \cite{Danzl2008} we could greatly improve the value for the binding energy of the rovibrational ground state $|5\!\!> = |v\!=\!0, J\!=\!0\!>$ by determining the binding energy of $|v\!=\!73>$ and using well-known data from conventional molecular spectroscopy \cite{Weickenmeier1985,Amiot2002}. Our measurement was limited by the calibration of our wavemeter, not allowing us to determine the number of the teeth of the frequency comb, and by the precision of the spectroscopy data. Searching for $|5\!\!>$ in dark state spectroscopy is now a straightforward task as only a range of about 0.002 wavenumbers needs to be scanned.
We do this by exciting the transitions from $|3\!\!>$ to $|4\!\!>$ with laser $L_3$ and from $|4\!\!>$ to $|5\!\!>$ with laser $L_4$ simultaneously. The intensity for $L_4$ is typically $5\cdot10^4$ mW/cm$^2$. As is well known, the two light fields create a molecule-molecule dark state. The molecules initially in $|3\!\!>$ are lost unless laser $L_4$ is on two-photon resonance, provided that the Rabi frequency $\Omega_4$ on the fourth transition is equal to or greater than $\Omega_3$, the Rabi frequency on the third transition. We look for the resonance condition with the rovibrational ground state $|v\!=\!0, J\!=\!0>$ for some of the excited levels that we found above. Table \ref{tab:1} lists the observed transition wavelengths. We check that we can identify the level with rotational quantum number $J\!=\!2$ as the rotational energy splitting is well known. Fig.\ref{fig:4} shows typical molecular dark resonances when we set $L_4$ on resonance and step the detuning $\Delta_3$ of $L_3$ near $1350$ nm.
From a three-level model matched to the data for the dark resonances, taking into account off-resonant excitations and laser line widths, we determine the molecular transition strengths as given by the normalized Rabi frequencies. One of the two-photon transitions appears to be a particularly good candidate for STIRAP ground state transfer. It involves the excited state level $|4\!\!>$ with vibrational number $v'=61$ of the $(A^1\Sigma_u^+-b^3\Pi_{0u})\; 0_{u}^{+}$ coupled state system. For the transition from $|3\!\!>$ to $|4\!\!>$ and from $|4\!\!>$ to $|5\!\!>$ the normalized Rabi frequencies are $\Omega_3\!=\!2\pi\!\times\!6$ kHz $ \sqrt{I/(\mathrm{mW/cm}^2)}$ and $\Omega_4\!=\!2\pi\!\times\!5$ kHz $ \sqrt{I/(\mathrm{mW/cm}^2)}$, respectively. These values carry an estimated error of 50\% as the laser beam parameters for $L_3$ and $L_4$ are not well determined. A comparison with a typical atomic transition strength of $\Omega_a\!=\!2\pi\!\times\!5$ MHz $ \sqrt{I/(\mathrm{mW/cm}^2)}$ giving $ |\Omega_3/\Omega_a|^2 \approx 10^{-6} $ and $ |\Omega_4/\Omega_a|^2 \approx 10^{-6} $ reflects the minuteness of the wave function overlap. Nevertheless, their value is sufficient for STIRAP as seen in our recent work \cite{Danzl2008}. Also, they are of similar magnitude. This facilitates STIRAP, for which the peak Rabi frequencies should be approximately equal for optimum performance.

\begin{table}[hbt]
\caption{Levels of the excited $0_u^+$ coupled state system in the region 9893 cm$^{-1}$ to 10091 cm$^{-1}$ above $X^1\Sigma_g^+$ $|v\!=\!0, J\!=\!0>$. The first column gives the coupled channel vibrational numbers of the individual levels. Levels marked with $^{\ast}$ have not been searched for and the level energies given are those determined from the coupled channels calculations. The column labeled 'C' gives the predominant contribution to the overall vibrational wave function, which is either predominantly $A^1\Sigma_u^+$ or predominantly $b^{3}\Pi_{0u}$, indicated by $A$ and $b$, respectively. The number in brackets gives the order within the two progressions of levels with either predominantly $A^{1}\Sigma_u^+$ or predominantly $b^3\Pi_{0u}$ character. Both the $|J\!=\!1>$ and the $|J\!=\!3>$ rotational levels were identified for all oberved excited state levels. The wavemeter accuracy gives a typical uncertainty in wavelength of $\pm0.002\,$nm, which translates into $\pm0.011\,$cm$^{-1}$ uncertainty in the value for the energy above $|v\!=\!0, J\!=\!0>$. The energy  relative to $X^1\Sigma_g^+$ $|v\!=\!0, J\!=\!0>$ of experimentally determined levels is based on the measured excitation wavelength from  $X^1\Sigma_g^+$ $|v\!=\!73, J\!=\!2>$ and the $X^1\Sigma_g^+$ $|v=73>$ level energy from Ref. \cite{Amiot2002}, which introduces an additional uncertainty of 0.001 cm$^{-1}$. Deexcitation wavelengths are obtained from dark resonance spectroscopy involving the respective intermediate excited state level and the rovibronic ground state $X^1\Sigma_g^+$ $|v\!=\!0, J\!=\!0>$. n. m.: not measured}
\label{tab:1}
\begin{tabular*}{0.48\textwidth}{rccp{0.09\textwidth}p{0.09\textwidth}p{0.09\textwidth}}
\hline\hline\noalign{\smallskip}
$v'$ & C & $J$ & Excitation wavelength from $X^1\Sigma_g^+$ $|v\!=\!73, J\!=\!2>$ [nm]& Energy above $X^1\Sigma_g^+$ $|v\!=\!0, J\!=\!0>$ [cm$^{-1}$] & De-excitation wavelength to $X^1\Sigma_g^+$ $|v\!=\!0, J\!=\!0>$ [nm]\\
\noalign{\smallskip}\hline\noalign{\smallskip}

57 & A (7)&1&1365.148 & 9893.002 & n. m.\\
57 & A (7)&3& 1365.131 & 9893.094 & n. m.\\
$^{\ast}$58 & b (50)&0& 1362.893 & 9905.126 & n. m.\\
$^{\ast}$59 & A (8)&0& 1357.748 & 9932.927 & n. m.\\
60 & b (51)&1&1357.091 & 9936.497 & n. m.\\
60 & b (51)&3&1357.071 & 9936.606 & n. m.\\
61 & b (52)&1&1351.367 & 9967.707 & 1003.240\\
61 & b (52)&3&1351.347 & 9967.816 & n. m.\\
$^{\ast}$62 & A (9)&0& 1350.388 & 9973.068 & n. m.\\
63 & b (53)&1&1345.725 & 9998.729 & 1000.128\\
63 & b (53)&3&1345.705 & 9998.839 & n. m.\\
$^{\ast}$64 & A (10)&0& 1343.082 & 10013.351 & n. m.\\
65 & b (54)&1&1340.162 & 10029.576 & 997.052\\
65 & b (54)&3&1340.143 & 10029.682 & n. m.\\
66 & A (11)&1& 1335.833 & 10053.759 & 994.653\\
66 & A (11)&3& 1335.816 & 10053.853 & n. m.\\
$^{\ast}$67 & b (55)&0& 1334.675 & 10060.249 & n. m.\\
68 & b (56)&1& 1329.257 & 10090.794 & 991.003\\
68 & b (56)&3& 1329.238 & 10090.902 & n. m.\\

\noalign{\smallskip}\hline\hline
\end{tabular*}
\end{table}

\section{Conclusion}
\label{sec:5}We observe several two-photon dark resonances that connect the intermediate rovibrational level $|v\!=\!73,  J\!=\!2>$ of the $X^1\Sigma_g^+$ ground state potential with the rovibrational ground state level $|v\!=\!0,  J\!=\!0>$. At least one of the two-photon transitions is sufficiently strong for implementing STIRAP to $|v\!=\!0,  J\!=\!0>$ in the quantum gas regime, paving the way for the realization of a BEC of ground state molecules. STIRAP can in principle be implemented in two ways, either in the form of two sequential two-photon STIRAP steps, or in the form of four-photon STIRAP \cite{Shore1991,Kuznetsova2008}.
An attractive strategy for the production of a BEC of ground state molecules relies on the addition of an optical lattice.
Starting from an atomic BEC, pairs of atoms at individual lattice
sites are produced in a superfluid-to-Mott-insulator transition \cite{Greiner2002}. These
pairs can then be very efficiently associated on a Feshbach resonance and
subsequently transfered to the rovibronic ground state
with STIRAP. The lattice has the advantage of shielding the molecules
against inelastic collisions during the association process and
subsequent state transfer. As proposed by Jaksch {\em et al.} \cite{Jaksch2002}, dynamical melting of the lattice should ideally result in the formation of a BEC of molecules in the rovibronic ground state in a Mott-insulator-to-superfluid-type transition.

\section{Acknowledgements}
We are indebted to R. Grimm for generous support and we thank E. Tiemann for valuable discussions and C. Amiot for providing the FTS data of LAC on Cs$_{2}$. We gratefully acknowledge funding
by the Austrian Ministry of Science and Research (BMWF) and the
Austrian Science Fund (FWF) in form of a START prize grant and by the European Science Foundation (ESF) in the framework of the EuroQUAM collective research project QuDipMol. R.H. acknowledges support by the European Union in form of a Marie-Curie International Incoming Fellowship (IIF). The work at Stony Brook was supported by the US NSF, under grant PHY0652459.

\end{document}